\documentclass[12pt,tightenlines,eqsecnum,floats,aps,amsmath,amssymb,nofootinbib,prd,showpacs]{revtex4}

\usepackage{setspace}
\usepackage{amsmath,amssymb,amsfonts,amsthm}
\usepackage{graphicx}
\usepackage{enumerate}

\newcommand{\sech}{\mathrm{sech\,}}

\begin{document}

\title{Unitary evolution of free massless fields in de Sitter space-time}

\author{Daniel  \surname{G\'omez Vergel}}
\email[]{dgvergel@iem.cfmac.csic.es} \affiliation{Instituto de
Estructura de la Materia, CSIC, Serrano 121, 28006 Madrid, Spain}
\author{Eduardo J. \surname{S. Villase\~nor}}
\email[]{ejsanche@math.uc3m.es} \affiliation{Grupo de
Modelizaci\'on y Simulaci\'on Num\'erica, Universidad Carlos III
de Madrid, Avda. de la Universidad 30, 28911 Legan\'es, Spain}
\affiliation{Instituto de Estructura de la Materia, CSIC, Serrano
123, 28006 Madrid, Spain}

\date{February 26, 2008}

\begin{abstract}
We consider the quantum dynamics of a minimally coupled massless scalar field in de
Sitter space-time. The classical evolution is represented by a
canonical transformation on the phase space for the field theory. By
studying the corresponding Bogoliubov transformations, we show that
the symplectic map that encodes the evolution between two instants
of time cannot be  unitarily implemented on any Fock space built
from a $SO(4)$-symmetric complex structure. We will show also that,
in contrast with some effectively lower dimensional examples arising
from Quantum General Relativity such as Gowdy models, it is
impossible to find a time dependent conformal redefinition of the
massless scalar field leading to a quantum unitary dynamics.

\end{abstract}


\maketitle

\section{Introduction}{\label{Intro}}

One of the most surprising features of any Fock quantization of a
linear symplectic dynamical system with infinitely many degrees of
freedom is the impossibility of defining the unitary quantum
counterpart of \textit{every} linear symplectic transformation
\cite{Shale}. In the special case of a field theory on a
four dimensional space-time background, this characteristic behavior of
the infinite dimensional systems is responsible for the generic
impossibility to make sense of the unitary quantum evolution operator from an
initial Cauchy surface to any final Cauchy surface, both for curved and flat background space-times \cite{Helfer:1996my}.
This is the case for scalar fields propagating in Minkowski space-time if the Cauchy
surfaces are not level surfaces of some Minkowskian time \cite{Torre:1998eq}.

\bigskip

The problem of the unitary implementability of symplectic
transformations lies also at the heart of many recent interesting
results concerning the quantization of some \textit{midi-superspace}
models \cite{Varadarajan1,Varadarajan2,Pierri:2000ri,Corichi:2002vy,Torre:2002xt,Corichi:2005jb,Corichi:2006xi,Corichi:2006zv,
Mena:2007,BarberoG.:2006zw,BarberoG.:2007,G.:2007rd}. Among these
symmetry reductions of general relativity, linearly polarized Gowdy models
\cite{Gowdy:1971jh,Gowdy:1973mu} have been the focus of intensive
study due to their cosmological interpretation and exact
solvability. After partial gauge fixing and
deparameterization\footnote{The deparameterization program  was done for the first
time in the context of the $\mathbb{T}^3$ Gowdy  model
\cite{Pierri:2000ri,Torre:2002xt,Corichi:2006xi}, and generalized
later for the $\mathbb{S}^{1}\times \mathbb{S}^{2}$ and
$\mathbb{S}^3$ cases  \cite{BarberoG.:2007}.}, the field-like
degrees of freedom of the Gowdy models admit a natural
interpretation as a massless scalar field $\phi$ propagating in a
globally hyperbolic $(1+2)$-background $(I_\Sigma\times \Sigma,
g^\Sigma_{ab})$. The scalar field dynamics has to satisfy an
additional symmetry condition with respect to a certain Killing
vector field of the background metric. The topology of the spatial
Cauchy surfaces  $\{t\}\times\Sigma$  is constrained to be
homeomorphic to $\mathbb{T}^2=\mathbb{S}^1\times \mathbb{S}^1$ (for
the so-called Gowdy $\mathbb{T}^3$ model) or $\mathbb{S}^2$ (for the
Gowdy $\mathbb{S}^{1}\times\mathbb{S}^{2}$ and $\mathbb{S}^3$
models). The time interval $I_\Sigma\subset \mathbb{R}$,
parameterized by a global coordinate $t$, is respectively $I_{\mathbb{T}^2}=(0,\infty)$
or $I_{\mathbb{S}^2}=(0,\pi)$. Finally, the space-time metric
$g_{ab}^\Sigma$ has a very simple form
\begin{equation}
g^\Sigma_{ab}=f_\Sigma^{-4}(t)\,\mathring{g}_{ab}^\Sigma
=f_\Sigma^{-4}(t)\big(-(\mathrm{d}t)_a(\mathrm{d}t)_b+\gamma^\Sigma_{ab}\big)\,,\label{Gowdy}
\end{equation}
where $\gamma^{\Sigma}_{ab}$ is the standard flat metric when
$\Sigma$ is the 2-torus, and the round metric when $\Sigma$ is the
2-sphere. The fuction $f_\Sigma$ is a time dependent conformal factor whose
explicit value is $f_{\mathbb{T}^2}(t)=1/\sqrt{t}$ or
$f_{\mathbb{S}^2}(t)=1/\sqrt{\sin t}$, depending on the spatial
topology.  At this point, it is worthwhile noting that the
background metric of the $\Sigma=\mathbb{S}^2$ models is somehow
similar to the ($1+3$)-dimensional de Sitter metric in the sense
that the metric (\ref{Gowdy}) is conformally equivalent to the
Einstein static ($1+2$)-universe. In all these cases, it is
impossible to find a Fock quantization using a complex structure
compatible with the \textit{spatial} symmetries of the Riemannian
metrics $\gamma^\Sigma_{ab}$, such that the quantum evolution (in
terms of the time coordinate $t$) of the massless field $\phi$ is
unitarily implementable\footnote{This was first shown for the
$\mathbb{T}^3$ model \cite{Corichi:2002vy,Mena:2007} and later on
seen to be true also in the  $\mathbb{S}^1\times \mathbb{S}^2$ and
$\mathbb{S}^3$ Gowdy topologies \cite{G.:2007rd}.}. However,
irrespective of $\Sigma$, it is always possible to avoid this
problem by performing a suitable time-dependent field redefinition
that preserves the spatial homogeneity and isotropy\footnote{Again,
these field redefinitions were firstly used to deal with the problem
of the unitary implementability of the quantum dynamics for the
$\mathbb{T}^3$ Gowdy model \cite{Corichi:2005jb} and later on
generalized --and interpreted on geometric grounds as conformal transformations--
for the remaining topologies \cite{G.:2007rd}.}. This is given by
$\phi(t,s)=f_\Sigma(t)\cdot\zeta(t,s)$ --here $(t,s)\in
I_\Sigma\times \Sigma$. When this is done, there exists a unique
(modulo unitary equivalence) Fock quantization for the new field
$\zeta$ such that the quantum time evolution can be unitarily
implemented \cite{Corichi:2006zv,G.:2007rd}.

\bigskip

Although the previous results were obtained in the restricted
context of the quantization of Gowdy models, one could expect that
the techniques described above have a wider range of applicability.
In particular, they could be used to identify --and possibly solve--
the problems associated with the failure of the unitary
implementation of time evolution in the Fock quantizations on
sufficiently symmetric curved backgrounds. In this respect, the most
ambitious program would be to formulate a theorem classifying the
class of space-times where a conformal redefinition of the fields
solves the problem of unitarity of evolution. However, before
starting such a general program, it is convenient to analyze
concrete models to acquire some familiarity with the different
possible behaviors and pathologies. In particular, it is important
to clarify if the dimensionality of the space-time plays a key role in
this process since, up to now,  only (1+2)-dimensional cases have
been analyzed. The main aim of this paper is precisely to test the techniques
developed in the context of Gowdy models in a (1+3)-dimensional
example. To this end, we will consider the class of Fock
quantizations for the \textit{minimally coupled massless scalar field}
propagating in the de Sitter space-time, not only owing to its
similarity to Gowdy models, but also due to its intrinsic interest.

\bigskip

The terms `mass' and `massless' deserve special attention when we are dealing with a curved background. This is so because the physical quantity called \textit{mass} can be associated both to the
Galilean or Poincare groups but it does not exist, by itself, as a conserved
quantity in a generic space-time. Nonetheless, it is possible to make sense of the mass
in de Sitter space by following an ambiguity free limit process in which the space-time becomes flat \cite{Fronsdal}.
In this context, the scalar field that we are going to consider in  this paper
is not really a massless field in the above sense  --in particular, its field equation is not conformally invariant--  but corresponds to the unitary irreducible representation associated with the zero-value of the Casimir operator (that is proportional to the Laplace-Bertrami operator) of the de Sitter group \cite{Gazeau:1999mi}. Although this scalar field is not massless, it arises in a natural form in the context of the massless spin-two fields in the de Sitter space and it is usually called the minimally coupled massless scalar field.

\bigskip

The problem of quantizing this system
has been profusely analyzed in the literature \cite{Gazeau:1999mi,Mottola:1984ar,Mottola:1985qt, Allen:1985ux, Allen:1987,  Pathinayake:1988au, Tolley:2001gg,Garriga}. The scalar quantum field $\hat{\phi}(x)$, understood
as an operator-valued distribution on certain Hilbert space
$\mathcal{H}$, must be a solution to the classical field equations (that we will discuss in section \ref{Scalar field on de Sitter background})
and satisfy certain reasonable physical conditions that can be
summarized as follows.

\begin{itemize}

\item[c1)] \textit{Microcausality.} The field commutator $[\hat{\phi}(x_1),\hat{\phi}(x_2)]$ must vanish if the points $x_1$ and $x_2$ of the de Sitter space-time are not causally connected.

\item[c2)] \emph{Covariance.} There exists a unitary representation $\hat{D}(g)$ of the de Sitter group
$O(1,4)$ such that the field is covariant, i.e.
$\hat{D}(g)\hat{\phi}(x)\hat{D}^{-1}(g)=\hat{\phi}(g\cdot x)$ for
all $g\in O(1,4)$, where we have denoted by $g\cdot x$ the action of
$g$ on the point $x$ of the de Sitter space.

\item[c3)] \textit{Invariance of the vacuum.} There exists a cyclic unit vector $|0\rangle\in \mathcal{H}$ which is invariant under the representation $\hat{D}$, i.e. $\hat{D}(g)|0\rangle =|0\rangle$, $\forall\,g\in O(1,4)$.

\item[c4)] \textit{Hadamard condition.} The Hadamard restriction demands that the two-point
functions
$\langle 0|\hat{\phi}(x_1)\hat{\phi}(x_2)|0\rangle$ have the right
short-distance singularity behavior as $x_1\rightarrow x_2$. This
condition allows us to define the expectation value of the
stress-energy operator in a completely satisfactory manner.

\end{itemize}

It can be shown that, once a non-divergent, complete set of
solutions to the equation of motion has been fixed, it is
impossible to construct a de Sitter-invariant Fock representation
for which the vacuum state is of Hadamard type
\cite{Allen:1985ux,Allen:1987}. The origin of the problem is the
existence of zero-modes, which is related to the fact that the wave
equation has constant solutions. Therefore, a fully covariant
quantization requires a type of representations of the canonical
commutation relations different from those used in any Fock-like
approach \cite{Gazeau:1999mi,Garriga}. The Gupta-Bleuer-type quantization scheme,
rigourously formulated in \cite{Gazeau:1999mi}, successfully
incorporated the conditions c1)-c4) stated above in a non-Fock
representation. However, even if it is possible to obtain a fully covariant quantization through non-Fock representations, we believe that it is necessary and interesting to probe if Fock quantizations based in
some suitable weakened version of covariance can provide a well
defined and unitary quantum dynamics in this system.  For instance, following  \cite{Allen:1987},
if one adopts a weaker version of covariance by restricting the condition c2) only to the subgroup
 $O(4)$ it is not
necessary to abandon the usual Fock quantization. Of course,
according to \cite{Helfer:1996my}, in those cases if the Fock vacua are assumed
to be of Hadamard type, it is expected that the symplectic transformation defined by the
time evolution cannot be, in general, unitarily implemented. Hence, if we want to have a unitary evolution, it is important to avoid restricting the discussion to the Hadamard case.

\bigskip

The differences of our treatment with respect to previous works available in literature can be
summarized as follows. We have not restricted ourselves to the de Sitter-invariant case but we have considered all the Fock quantizations for which the one-particle Hilbert space is obtained from a
$SO(4)$-invariant complex structure. Furthermore, we do not require  the states
to be of Hadamard type (a physically well justified hypothesis that
nevertheless makes the unitary implementability of dynamics
difficult \cite{Helfer:1996my}). In addition, in order to study the unitary implementability of the quantum evolution, we explore all the possible homogeneous reparameterizations of the minimally coupled scalar field.

\bigskip

Before going further, it is important to discuss to what extent the
treatment for the Gowdy models introduced above can be followed in
the de Sitter case. For the vacuum Gowdy models the background metric is
fiducial (i.e. it is a by-product of the deparameterization process)
and the massless scalar field has only a geometric interpretation,
but not a physical one. Therefore, we are free to perform conformal
redefinitions of the form
$$
g_{ab}^\Sigma\mapsto \mathring{g}^\Sigma_{ab}=f^4_\Sigma g_{ab}^\Sigma\,,
\quad \phi\mapsto \zeta =f^{-1}_\Sigma \phi\,,
$$
that modify both the background metric and the scalar field.
However, in the context of quantum field theory in curved
space-times, both the matter fields and the background metric have a
clear physical interpretation and are chosen to describe a concrete
physical situation. Hence, \emph{a priori}, the class of conformal
transformations considered in order to deal with the Gowdy models are in this
case more difficult to justify. The only acceptable reason for
modifying the background and the matter field would be to avoid some
problems associated with the original theory, such as the failure of
the unitary implementation of the quantum dynamics.

\bigskip

The paper is organized in the following way. After this
introduction, we will review in section \ref{de Sitter space-time}
the properties of the de Sitter space-time that are used in the body
of the paper. We will discuss in section \ref{Scalar field on de
Sitter background} the different formulations describing a minimally coupled massless
scalar field propagating in de Sitter space-time that are available
after performing a conformal re-scaling of the massless field. In
section \ref{Quantum Hilbert space} we will explicitly characterize
all the Fock quantizations of the massless scalar field that respect
the $SO(4)$ symmetry of the spatial slices of the de Sitter
background. The main result of the paper is presented in section
\ref{Classical dynamics and unitary quantum evolution} where we will
show that, irrespective of the conformal factor of the massless
scalar, it is impossible to have a quantum unitary evolution
operator. Notice that this result affects the two-parameter family
of $O(4)$-invariant Hadamard representations characterized in
\cite{Allen:1987}. We will end the paper with our conclusions in
section \ref{Conclusions and comments}.

\section{de Sitter space-time}{\label{de Sitter space-time}}

Space-time metrics $g_{ab}$ of constant curvature are locally
characterized by the vanishing of their Weyl curvature. In $(1+3)$
dimensions this implies
\begin{eqnarray}
R_{ab}-\frac{1}{4}Rg_{ab}=0\,,\quad R \textrm{ constant}. \label{Eins}
\end{eqnarray}
The  solution to (\ref{Eins}) that corresponds to
 positive constant curvature $R>0$ is the so-called \emph{de Sitter
space-time} $\mathrm{dS}(R)$.  This one-parameter family of
space-times can be explicitly constructed as the warped product
$\mathbb{R}\times_{\kappa_R} \mathbb{S}^3$ of the real line, with
metric element $-\mathrm{d}t^2$, and the round 3-sphere by
considering the warping functions\footnote{We will choose units such
that $c=\hbar=1$.}
$$\kappa_R(t)=\frac{12}{R}\cosh^{2}\left(\sqrt{\frac{R}{12}}\,
t\right).$$ In other words
\begin{equation}\label{dSmetric}
g_{ab}=-(\mathrm{d}t)_{a}(\mathrm{d}t)_{b}+H^{-2}\cosh^{2}(Ht)\gamma_{ab}\,,
\end{equation}
where $H:=\sqrt{R/12}$ is the Hubble constant and $\gamma_{ab}$ is
the pullback to $\mathrm{dS}(R)$ of the round metric of the unit
3-sphere by means of the canonical projection $\mathbb{R}\times
\mathbb{S}^3\rightarrow \mathbb{S}^3$. The spatial slices of
constant time coordinate $t$ foliate $\mathrm{dS}(R)$ by a
one-parameter family of Cauchy surfaces all of them diffeomorphic to
$\mathbb{S}^3$. Their geodesic normals contract monotonically to a
minimum spatial separation at $t=0$, and then re-expand
exponentially to infinity. In view of (\ref{Eins}), $\mathrm{dS}(R)$
can be regarded as a solution of vacuum Einstein's equation with a
positive cosmological constant $\Lambda=R/4>0$ which sets the
expansion rate to $H\propto\sqrt{\Lambda}$. In this context, if the
observed acceleration of our universe can be explained in terms of a
cosmological constant, its evolution may be described by a de Sitter
solution at the infinite future.

The structure of the conformal infinity of $\mathrm{dS}(R)$ can be
easily obtained from (\ref{dSmetric}) by redefining the global time
coordinate $t\in \mathbb{R}$ in terms of a  new global time
coordinate $T\in (-\pi/2,\pi/2)$ through
\begin{equation}
T=2\arctan\left(\tanh(Ht/2)\right),\quad -\pi/2< T<
\pi/2\,.\label{Ttime}
\end{equation}
In terms of this coordinate it is clear that
\begin{equation}
g_{ab}=H^{-2}\cosh^{2}(Ht)\,\bar{g}_{ab}\,,\quad\textrm{ where }\,\,
\bar{g}_{ab}:=-(\mathrm{d}T)_{a}(\mathrm{d}T)_{b}+\gamma_{ab}\,.\label{dS_conformal}
\end{equation}
Therefore, denoting by
$E:=(\mathbb{R}\times\mathbb{S}^{3},\bar{g}_{ab})$ the Einstein
static universe, with $-\infty<T<\infty$, the de Sitter space is
conformal to the submanifold
$((-\pi/2,\pi/2)\times\mathbb{S}^{3},\bar{g}_{ab})\subset E$. Then,
the surfaces $T=\pm\pi/2$ act as future and past space-like
infinities for time-like and null lines in de Sitter space-time. Due
to this fact, and in contrast to Minkowski space, both particle and
event horizons exist in the de Sitter space-time for geodesic
families of observers \cite{HawkingEllis}.

In the following we will adimensionalize the time coordinate by
changing $t\mapsto Ht$, and choose units such that $H=1$. Then we
can write
\begin{equation}
g_{ab}=-(\mathrm{d}t)_{a}(\mathrm{d}t)_{b}+\cosh^2t\,\gamma_{ab}\,.\label{dSadim}
\end{equation}

\section{Scalar field on de Sitter background}{\label{Scalar field on de Sitter background}}

In this paper we will focus on the dynamics of free minimally coupled massless scalar
fields propagating in a de Sitter background. To this end, let us
consider the standard classical action
\begin{equation}
S(\phi)=-\frac{1}{2}\int_{[t_0,t_1]\times
\mathbb{S}^{3}}|g|^{1/2}g^{ab}(\mathrm{d}\phi)_{a}(\mathrm{d}\phi)_{b}\,\label{massless_action}
\end{equation}
for a smooth free massless scalar field $\phi\in
C^\infty(\mathbb{R}\times \mathbb{S}^3;\mathbb{R})$ evolving from
$t_0$ to $t_1$ in a de Sitter space-time $(\mathbb{R}\times
\mathbb{S}^3,g_{ab})$, where $g_{ab}$ is given by (\ref{dSadim}).
The massless field $\phi$ satisfies then the linear wave equation
\begin{equation}
g^{ab}\nabla_a\nabla_b \phi=0\,, \label{ecuacion}
\end{equation}
where $\nabla_a$ is the Levi-Civita connection associated to
$g_{ab}$. It is well known \cite{WaldGR} that equation
(\ref{ecuacion}) has a well possed initial value formulation in
terms of the Cauchy data defined on some level surface of the
coordinate field $t$. In the following subsections we will discuss
certain alternative descriptions of this system and analyze in
detail the structure of its covariant and canonical phase space
formulations.

\subsection{Conformal field redefinitions}

As we have mentioned in the introduction, one of the main lessons
that we have learned from the study of the quantization of the Gowdy
models is that, in order to have a well defined unitary quantum
dynamics, it is imperative to consider a new auxiliary field $\zeta$
related to the original one in terms of a conformal transformation
\cite{BarberoG.:2007}. In  the case of a massless scalar field
propagating in the de Sitter space-time, we will restrict ourselves
to spatially homogeneous field redefinitions of the form
\begin{eqnarray}
\phi(t,s)=f(t)\zeta(t,s)\,,\quad  (t,s)\in\mathbb{R}\times\mathbb{S}^{3}\,, \label{redefinicion}
\end{eqnarray}
with $f(t)>0$ a (fixed) smooth, positive definite, real-valued
function that we will try to fix by demanding that the quantum
theory be well behaved.

The classical field dynamics for $\zeta$ can be derived from the
variational principle
\begin{eqnarray}
s(\zeta)&:=&S(\phi)=S(f\zeta)\label{action_0}
\\
&=&-\frac{1}{2}\int_{[t_0,t_1]\times \mathbb{S}^3} |\mathring{g}|^{1/2}
\mathring{g}^{ab}\Big( (\mathrm{d}\zeta)_a(\mathrm{d}\zeta)_b
+2(\mathrm{d} \log f)_a\zeta (\mathrm{d}\zeta)_b + (\mathrm{d}\log
f)_a(\mathrm{d}\log f)_b \zeta^2   \Big)\nonumber
\end{eqnarray}
that can be easily obtained by plugging (\ref{redefinicion}) in
(\ref{massless_action}). Here $\mathring{g}_{ab}:=f^2\,g_{ab}$ is a
space-time metric on $\mathbb{R}\times \mathbb{S}^3$ conformal to
the de Sitter one. Notice that the field $\zeta$ can be interpreted
as a scalar field propagating on the metric background defined by
$\mathring{g}_{ab}$ and coupled, in a non-standard way, to the
time-dependent  potential $\log f$. The  massless field equation (\ref{ecuacion}) for $\phi$ can be  equivalently rewritten in terms of $\zeta$ in the form
\begin{equation}
\mathring{g}^{ab}\mathring{\nabla}_a\mathring{\nabla}_b\zeta+\Big(\mathring{g}^{ab}\mathring{\nabla}_a\mathring{\nabla}_bf -\mathring{g}^{ab}(\mathrm{d}\log f)_a(\mathrm{d}\log f)_b\Big)\zeta=0\,,\label{ecuacion_zeta}
\end{equation}
where $\mathring{\nabla}_a$ is the Levi-Civita connection associated to
$\mathring{g}_{ab}=f^2\,g_{ab}$. Equation (\ref{ecuacion_zeta}) admites the interpretation of a Klein-Gordon like equation for the field $\zeta$ --in the background $\mathring{g}_{ab}$-- with a (generically) nonconstant `mass term'.   There are at least two special cases that deserve special attention.  First, if we take $f(t)=1$ we recover the original massless scalar field propagating in de Sitter
space-time. On the other hand, in view of (\ref{dS_conformal}), if
we choose $f(t)=1/\cosh t$  the metric $\mathring{g}_{ab}$ becomes
the metric of the Einstein static universe.

The action (\ref{action_0}) can be equivalently written as
\begin{eqnarray}
s(\zeta)&=&\frac{1}{2}\int_{t_0}^{t_1} \mathrm{d}t
\int_{\mathbb{S}^{3}}
|\gamma|^{1/2}\bigg(f^{2}\cosh^{3}(t)\dot{\zeta}^{2}+2f\dot{f}\cosh^{3}(t)\zeta\dot{\zeta}
\label{action}\\
& & \hspace{5cm}+f^{2}\cosh(t)\zeta \Delta_{\mathbb{S}^3} \zeta
+\dot{f}^{2}\cosh^{3}(t)\zeta^{2}\bigg)\,,\nonumber
\end{eqnarray}
where the dot denotes the Lie derivative along $(\partial/\partial
t)^{a}$ and $\Delta_{\mathbb{S}^3}$ is the Laplace-Beltrami  operator on the
round unit 3-sphere.

\subsection{Covariant phase space}

The space $\mathcal{S}$ of smooth real solutions to the equation of
motion derived from the action (\ref{action}) has the structure of
an infinite-dimensional real vector space. Any $\zeta\in
\mathcal{S}$ is a real smooth function on $\mathbb{R}\times
\mathbb{S}^3$ that satisfies the wave equation
\begin{equation}\label{symplectic}
\ddot{\zeta}+\left(3\tanh(t)+2\frac{\dot{f}(t)}{f(t)}\right)\dot{\zeta}
+\left(3\tanh(t)\frac{\dot{f}(t)}{f(t)}+\frac{\ddot{f}(t)}{f(t)}
-\frac{\Delta_{\mathbb{S}^{3}}}{\cosh^{2}(t)}\right)\zeta=0\,.
\end{equation}
The variational principle (\ref{action}) gives rise also to a natural
(weakly) symplectic structure $\Omega$ on $\mathcal{S}$ defined by
\begin{equation}\label{OmegaS}
\Omega(\zeta_1,\zeta_2):=
f^{2}(t)\cosh^{3}(t)\int_{\mathbb{S}^{3}}|\gamma|^{1/2}\imath_{t}^{*}(\zeta_1\dot{\zeta}_2-\zeta_2\dot{\zeta}_1)\,,
\end{equation}
where $\imath_t$ denotes the standard inclusion
$\imath_t:\mathbb{S}^3\rightarrow \mathbb{R}\times \mathbb{S}^3$,
$\imath_t(s)=(t,s)$. It is easy to show that $\Omega$ does not
depend upon the choice of time $t$ used to define the embedding
$\imath_{t}(\mathbb{S}^{3})\subset \mathbb{R}\times \mathbb{S}^3$.
The (infinite dimensional) linear symplectic space
$\Gamma=(\mathcal{S},\Omega)$ is called the \emph{covariant} phase
space of the system.

\subsection{Canonical phase space}

Let
$\Upsilon=(\mathcal{P},\omega)$ be the \emph{canonical}
phase space of smooth initial Cauchy data on the 3-sphere
$$
\mathcal{P}:=\{(Q,P)\,|\,Q,P\in
C^{\infty}(\mathbb{S}^{3};\mathbb{R})\}\,,
$$
endowed with the standard
(weakly) symplectic structure
\begin{equation}\label{omega}
\omega((Q_1,P_1),(Q_2,P_2)):=\int_{\mathbb{S}^{3}}|\gamma|^{1/2}(Q_{1}P_{2}-Q_{2}P_{1})\,.
\end{equation}
Once a time $t$ has been specified, it is possible to construct a
symplectomorphism between the spaces $\Gamma$ and $\Upsilon$. The
isomorphism $\mathfrak{J}_{t}:\Gamma\rightarrow\Upsilon$, that takes
a solution $\zeta\in\Gamma$ and finds the Cauchy data induced on
$\mathbb{S}^{3}$ by virtue of the embedding
$\imath_{t}:\mathbb{S}^{3}\rightarrow\mathbb{R}\times \mathbb{S}^3$,
\begin{eqnarray}
& & \mathfrak{J}_{t}:\Gamma\rightarrow\Upsilon\,,\quad \zeta\mapsto (Q,P)=\mathfrak{J}_{t}(\zeta)\,,\nonumber\\
\mathrm{where } & & Q=\imath_{t}^{*}\zeta\,,\quad
P=f^{2}(t)\cosh^{3}(t)\imath_{t}^{*}\dot{\zeta}+f(t)\dot{f}(t)\cosh^{3}(t)\imath_{t}^{*}\zeta\,,\label{isomor}
\end{eqnarray}
is  (irrespective of the value of $t$) a symplectic transformation
from $\Gamma$ to $\Upsilon$, i.e.
$\Omega=\mathfrak{J}_{t}^{*}\omega$. The description of the
classical dynamics in $\Upsilon$ is done in the form a
non-autonomous Hamiltonian system
$(\mathbb{R}\times\Upsilon,\omega,H(t))$, with time-dependent
Hamiltonian $H(t):\Upsilon\rightarrow\mathbb{R}$ given by the
quadratic form
\begin{equation}\label{hamiltonian}
H(Q,P;t)=\frac{1}{2}\int_{\mathbb{S}^{3}}|\gamma|^{1/2}\left(\frac{P^{2}}{f^{2}(t)\cosh^{3}(t)}-2
\frac{\dot{f}(t)}{f(t)}PQ-f^{2}(t)\cosh(t)\, Q\Delta_{\mathbb{S}^3}Q\right).
\end{equation}
Notice that when $\dot{f}\neq 0$ the Hamiltonian (\ref{hamiltonian}) is indefinite.

It is important to point out that it is only a matter of convenience
whether we describe dynamics using the covariant or the canonical
phase space. For the most part of this paper, we will study the
dynamical aspects of the system by using the covariant phase space.
In particular, we will discuss in some detail in section
\ref{Classical dynamics and unitary quantum evolution} the classical
evolution and its quantum counterpart.

\subsection{Mode decomposition}

Any vector $\zeta\in\Gamma$ is a smooth function on
$\mathbb{R}\times \mathbb{S}^3$. Then, for each fixed  value of
$t\in \mathbb{R}$, $\imath_t^*\zeta \in
C^\infty(\mathbb{S}^3;\mathbb{R})$. It is well known that any smooth
function on $\mathbb{S}^3$ can be written in the form
\begin{equation}
\imath_t^*\zeta(s)=\zeta(t,s)=\sum_{k=0}^{\infty}\sum_{l=0}^{k}\sum_{m=-l}^{l} A_{kl m}(t) Y_{kl m} (s)\,.\label{modos}
\end{equation}
Here $Y_{klm}$ denote the standard spherical harmonics on
$\mathbb{S}^3$ \cite{Lehoucq}. They  are eigenvectors of the
Laplacian, $\Delta_{\mathbb{S}^{3}}Y_{klm}=-k(k+2)Y_{klm}$, and
verify the $L^2(\mathbb{S}^3)$-orthogonality condition ($\delta$
denotes the Kronecker delta)
\begin{equation}
\int_{\mathbb{S}^{3}}|\gamma|^{1/2}\bar{Y}_{k_1l_1m_1}Y_{k_2l_2m_2}=\delta(k_1,k_2)\delta(l_1,l_2)\delta(m_1,m_2)\,.
\end{equation}
They also satisfy\footnote{The bar denotes complex conjugation.}
$\bar{Y}_{klm}=Y_{kl-m}$\,. The complex coefficients $A_{kl m}(t)$
appearing in (\ref{modos}) are defined in terms of $\zeta$ through
\begin{equation}
A_{kl m}(t)=\int_{\mathbb{S}^3}|\gamma|^{1/2} \bar{Y}_{kl m}\, \imath^*_t\zeta\,.\label{A}
\end{equation}
From (\ref{A}) it is clear that $A_{kl m}(t)$ are smooth functions
on $t\in \mathbb{R}$ that satisfy
\begin{eqnarray}
& & \bar{A}_{kl m}(t) =A_{kl -m}(t)\,,\label{real_cond}\\
& & \lim_{k\rightarrow \infty}\frac{1}{k^p}\frac{\mathrm{d}^n A_{kl
m}}{\mathrm{d} t^n}(t)=0\,,  \quad \forall\,p,n\in
\mathbb{N}_0\,,\quad \forall\,t\in \mathbb{R}.\label{cond_serie}
\end{eqnarray}
The condition (\ref{real_cond}) comes from the reality of the field
$\zeta$. On the other hand, condition (\ref{cond_serie}) is derived
from the fact that, for any fixed value of $t$,
$\partial^n_t\zeta(t,\cdot)$ is a smooth function  on
$\mathbb{S}^3$. Then, it is clear that any $\zeta\in \Gamma$ can be
expanded in the form
\begin{equation}
\zeta(t,s)=\sum_{k=0}^{\infty}\sum_{l=0}^{k}\sum_{m=-l}^{l}\Big(a_{klm}y_{k}(t)Y_{klm}(s)
+\overline{a_{klm}y_{k}(t)Y_{klm}(s)}\Big)\,,\quad  (t,s)\in\mathbb{R}\times\mathbb{S}^{3}\,,\label{modes}
\end{equation}
where the complex smooth functions $y_{k}$ are solutions to the
equation
\begin{equation}\label{equaty}
\ddot{y}_{k}+\left(3\tanh(t)+2\frac{\dot{f}(t)}{f(t)}\right)\dot{y}_{k}
+\left(3\tanh(t)\frac{\dot{f}(t)}{f(t)}+\frac{\ddot{f}(t)}{f(t)}+\frac{k(k+2)}{\cosh^{2}(t)}\right)y_{k}=0\,,
\end{equation}
whose real and imaginary parts are taken to be linearly independent,
and the complex coefficients  $a_{kl m}$ satisfy\footnote{Form the
point of view of the classical theory, these conditions are needed
to guarantee the smoothness of the solutions to the field equations. However, we do not need to know them in detail to discuss the quantized the model. In fact they will be relaxed to the milder condition
$\sum_{klm}|a_{klm}|^2<\infty$ when we introduce the one-particle Hilbert space. }
$$
\lim_{k\rightarrow
\infty}\frac{1}{k^p}\left(a_{klm}\frac{\mathrm{d}^n y_k}{\mathrm{d}
t^n}(t)+\bar{a}_{kl-m}\frac{\mathrm{d}^n \bar{y}_k}{\mathrm{d}
t^n}(t)\right)=0\,,\quad \forall\,p,n\in \mathbb{N}_0\,,\quad
\forall\,t\in \mathbb{R}.
$$
By explicitly decomposing $y_{k}=u_{k}+iv_{k}$, with $u_{k}$ and $v_{k}$ real,  the Wronskian of $u_k$ and $v_k$ can be written in terms of the Wronskian of $y_k$ and $\bar{y}_k$
$$2iW_{k}(t;u_k,v_k):=2i\Big(\dot{v}_{k}(t)u_{k}(t)-\dot{u}_{k}(t)v_{k}(t)\Big)=\bar{y}_{k}(t)\dot{y}_{k}(t)
-\dot{\bar{y}}_{k}(t)y_{k}(t)\,.$$
By virtue of (\ref{equaty}), $W_k$
verifies the differential equation
\begin{equation}
\dot{W}_{k}+\left(3\tanh(t)+2\frac{\dot{f}(t)}{f(t)}\right)W_{k}=0\,.
\end{equation}
It is then possible to write
$$W_{k}(t;u_k,v_k)=f^{-2}(t)\cosh^{-3}(t)W_{k}^{0}\,,$$
with $W_{k}^{0}\in\mathbb{R}$. We will normalize the pairs
$(u_{k},v_{k})$ by imposing
\begin{equation}
W_{k}(t;u_k,v_k)=\frac{1}{2}f^{-2}(t)\cosh^{-3}(t)\,,\quad \forall\,
k\in\mathbb{N}_{0}\,.\label{normalizacion}
\end{equation}
It could appear that this condition is rather arbitrary but, as we
will see, it is convenient to make this choice to ensure that the
set of complex solutions
$$\{\phi_{klm}=y_kY_{kl m}\,|\, k\in\mathbb{N}_0, l\in\{0,\dots,k\},\, m\in\{-l,\dots,l\}\}$$
is an orthonormal basis of the one-particle Hilbert space used to
construct the physical Fock space of the system. In concrete, a
possible election satisfying (\ref{normalizacion}) is given by
\begin{eqnarray}
u_{0k}(t)&:=&\frac{2^{(k+1)}}{\sqrt{3}f(t)}\mathrm{e}^{-kt}\cosh^{k}\!t\,\, {_{2}}F_{1}(k+3/2,k;-1/2;-\mathrm{e}^{-2t})\,,\label{u0v0}\\
v_{0k}(t)&:=&-\frac{2^{(k+1)}}{\sqrt{3}f(t)}\mathrm{e}^{-(k+3)t}\cosh^{k}\!t\,\, {_{2}}F_{1}(k+3/2,k+3;5/2
;-\mathrm{e}^{-2t})\,,\nonumber
\end{eqnarray}
where ${_{2}}F_{1}(a,b;c;z)$ are hypergeometric functions
\cite{WhittakerWatson}. In this way, we get a very convenient
expression for the symplectic structure (\ref{OmegaS}) that will be
our starting point for the quantization of the system
\begin{eqnarray}
\Omega(\zeta_1,\zeta_2)&
=&i\sum_{k=0}^{\infty}\sum_{l=0}^{k}\sum_{m=-l}^{l}(\bar{a}_{klm}^{(1)}a_{klm}^{(2)}-\bar{a}_{klm}^{(2)}a_{klm}^{(1)})\,,
\\
\mathrm{where} \quad \zeta_\alpha&=&\sum_{k=0}^{\infty}\sum_{l=0}^{k}\sum_{m=-l}^{l}\Big(a^{(\alpha)}_{klm}\phi_{klm}
+\bar{a}_{klm}^{(\alpha)}\bar{\phi}_{klm}\Big)\,,\quad \alpha=1,2\,.
\end{eqnarray}

\section{Quantum Hilbert space}{\label{Quantum Hilbert space}}

We will review in this section the Fock quantization technique based
on the covariant phase space of the model. It is well known
\cite{Wald} that  for a system of a finite number of uncoupled
quantum harmonic oscillators this procedure provides a quantum
theory unitarily equivalent to the usual tensor product of
one-particle Hilbert spaces. However, for the case of a system of
infinitely many uncoupled quantum harmonic oscillators, the tensor
product of the infinite one particle Hilbert spaces gives rise to a
non-separable Hilbert space, as well as reducible representations of
the canonical commutation relations. For these reasons the Fock quantization
introduced below provides a better  approach to deal with the
infinitely many degrees of freedom present in our model.

\subsection{One particle Hilbert space and its Fock space}

Let $\mathcal{S}_{\mathbb{C}}$ be the
complexification of the space of solutions $\mathcal{S}$ to the
equation of motion (\ref{symplectic}). Every vector
$Z\in\mathcal{S}_{\mathbb{C}}$ can be expressed as
\begin{equation}
Z(t,s)=\sum_{k=0}^{\infty}\sum_{l=0}^{k}\sum_{m=-l}^{l}\left(a_{klm}\phi_{klm}(t,s)+b_{klm}\bar{\phi}_{klm}(t,s)\right)\,,\quad
a_{klm}\,,\, b_{klm}\in\mathbb{C}\,.
\end{equation}
Let us extend $\Omega$ to $\mathcal{S}_{\mathbb{C}}$ by complex
linearity in each variable
\begin{equation}
\Omega_{\mathbb{C}}(Z_1,Z_2)=i\sum_{k=0}^{\infty}\sum_{l=0}^{k}\sum_{m=-l}^{l}\left(b_{klm}^{(1)}a_{klm}^{(2)}-b_{klm}^{(2)}a_{klm}^{(1)}\right),
\end{equation}
and consider the sesquilinear map $\langle\cdot,\cdot\rangle:\mathcal{S}_{\mathbb{C}}\times\mathcal{S}_{\mathbb{C}}\rightarrow\mathbb{R}$
defined by
\begin{equation}
\langle
Z_1,Z_2\rangle:=-i\Omega_{\mathbb{C}}(\bar{Z}_1,Z_2)=\sum_{k=0}^{\infty}\sum_{l=0}^{k}\sum_{m=-l}^{l}\left(\bar{a}_{klm}^{(1)}a_{klm}^{(2)}-\bar{b}_{klm}^{(1)}b_{klm}^{(2)}\right).
\end{equation}
This map satisfies all the properties of an inner product on
$\mathcal{S}_{\mathbb{C}}$ except that it fails to be positive
definite. However by considering the Lagrangian subspaces
\begin{eqnarray}
\mathbf{P}&:=&\big\{Z\in\mathcal{S}_{\mathbb{C}}\,:\,
Z=\sum_{k=0}^{\infty}\sum_{l=0}^{k}\sum_{m=-l}^{l}a_{klm}\phi_{klm}\big\}\,,\label{P}\\
\bar{\mathbf{P}}&:=&\big\{Z\in\mathcal{S}_{\mathbb{C}}\,:\,Z
=\sum_{k=0}^{\infty}\sum_{l=0}^{k}\sum_{m=-l}^{l}b_{klm}\bar{\phi}_{klm}\big\}\,,\label{bar_P}
\end{eqnarray}
it is possible to decompose
$\mathcal{S}_{\mathbb{C}}=\mathbf{P}\oplus \bar{\mathbf{P}}$ thus
obtaining a positive definite restriction
$\langle\cdot,\cdot\rangle|_{\mathbf{P}}$. The separable and
infinite-dimensional \textit{one-particle Hilbert space}
$\mathcal{H}_{\mathbf{P}}\cong\ell_2$ is obtained by Cauchy
completion
\begin{eqnarray*}
\mathcal{H}_{\mathbf{P}}
&:=&\overline{(\mathbf{P},\langle\cdot,\cdot\rangle|_{\mathbf{P}})}^{\langle\cdot,\cdot\rangle|_{\mathbf{P}}}=\big\{Z\,:\,
Z=\sum_{klm}a_{klm}\phi_{klm}\,,\, a_{klm}\in \mathbb{C}\,,\,
\sum_{klm}|a_{klm}|^2<\infty\big\}\,.
\end{eqnarray*}
The set $\{\phi_{klm}\}$ becomes a countable orthonormal basis of
$\mathcal{H}_{\mathbf{P}}$. The Hilbert space of the quantum field
theory $\mathcal{F}_{s}(\mathcal{H}_{\mathbf{P}})$ is then the
symmetric Fock space
\begin{equation}
\mathcal{F}_{s}(\mathcal{H}_{\mathbf{P}}):=\bigoplus_{n=0}^{\infty}\mathcal{H}^{\otimes_s^n}_{\mathbf{P}}\,,
\end{equation}
where $\mathcal{H}^{\otimes_s^n}_{\mathbf{P}}$ denotes the Hilbert space
of all $n$-th rank symmetric tensors over $\mathcal{H}_{\mathbf{P}}$.

\bigskip

Notice that every choice (\ref{P})-(\ref{bar_P}) of the Lagrangian
subspaces $\mathbf{P}$ and $\bar{\mathbf{P}}$ corresponds to the
specification of a complex structure $J_{\mathbf{P}}$ on the space
of solutions $\mathcal{S}$. Indeed, due to the fact that
$\mathbf{P}\cap\bar{\mathbf{P}}=\{0\}$, it follows that
$\mathcal{S}_{\mathbb{C}}=\mathbf{P}\oplus\bar{\mathbf{P}}$, and
hence any vector $\zeta\in\mathcal{S}$ can be uniquely decomposed as
$\zeta=Z+\bar{Z}$, with $Z\in \mathbf{P}$,
$\bar{Z}\in\bar{\mathbf{P}}$. Then, given $\mathbf{P}$ and
$\bar{\mathbf{P}}$, we can define the complex structure
$J_{\mathbf{P}}:\mathcal{S}\rightarrow\mathcal{S}$ by
$J_{\mathbf{P}}\zeta:=iZ-i\bar{Z}$. This map is a linear canonical
transformation on $\Gamma=(\mathcal{S},\Omega)$ --i.e.
$J_{\mathbf{P}}$ on $\mathcal{S}$ is compatible with $\Omega$-- such
that $J_{\mathbf{P}}^{2}=J_{\mathbf{P}}\circ
J_{\mathbf{P}}=-\mathrm{Id}_{\mathcal{S}}$ and the formula
\begin{equation}
\mu_{J_{\mathbf{P}}}(\zeta_1,\zeta_2):=\frac{1}{2}\Omega(J_{\mathbf{P}}\zeta_1,\zeta_2)
\end{equation}
defines a positive definite bilinear symmetric form on
$\mathcal{S}$. We conclude then that the sesquilinear map
\begin{equation}\label{innerprod}
\langle
\zeta_1,\zeta_2\rangle_{J_{\mathbf{P}}}:=\mu_{J_{\mathbf{P}}}(\zeta_1,\zeta_2)-\frac{i}{2}\Omega(\zeta_1,\zeta_2)
\end{equation}
is an inner product on $\mathcal{S}_{J_{\mathbf{P}}}$
\cite{Woodhouse}. Here $\mathcal{S}_{J_{\mathbf{P}}}$ is the complex
vector space $\cal S$ where, given any $\zeta\in\mathcal{S}$, the
product by complex scalars $\mathbb{C}\ni z=x+iy$,
$x,y\in\mathbb{R}$, is defined by the rule
$z\cdot\zeta:=x\zeta+yJ_{\mathbf{P}}\zeta$. In this context, the
one-particle Hilbert space  $\mathcal{H}_{\mathbf{P}}$ is given by
the Cauchy completion of the Euclidean space
$(S_{J_{\mathbf{P}}},\langle\cdot,\cdot\rangle_{J_{\mathbf{P}}})$.
It is straightforward to check that the Cauchy completions of
$(\mathbf{P},\langle\cdot,\cdot\rangle)$ and
$(S_{J_{\mathbf{P}}},\langle\cdot,\cdot\rangle_{J_{\mathbf{P}}})$
are isomorphic: the $\mathbb{C}$-linear map
$\kappa:S_{J_\mathbf{P}}\rightarrow \mathbf{P}$ such that
$\kappa(\zeta)=Z$, $\zeta\in\mathcal{S}_{J_\mathbf{P}}$, $Z\in
\mathbf{P}$, defines a unitary transformation of
$\mathcal{S}_{J_\mathbf{P}}$ onto $\mathbf{P}$, i.e.,
$\langle\zeta_1,\zeta_2\rangle_{J_\mathbf{P}}=\langle\kappa(\zeta_1),\kappa(\zeta_2)\rangle=\langle
Z_1,Z_2\rangle$,
$\forall\,\zeta_1,\zeta_2\in\mathcal{S}_{J_{\mathbf{P}}}$.

\subsection{Complex structures and mode decomposition}{\label{Complex structures and mode decomposition}}

In practice, the definition of the complex structure
$J_{\mathbf{P}}$ is complete once we are given a family of complex
functions $\{y_k\}$ satisfying (\ref{equaty}) and
(\ref{normalizacion}). In this case, we can construct an orthonormal
basis $\{\phi_{klm}=y_kY_{klm}\}$ of the one particle Hilbert space
$\mathcal{H}_\mathbf{P}$ and define $J_{\mathbf{P}}$ by imposing
that the complex structure is diagonalized in
$\mathcal{S}_\mathbb{C}$
\begin{equation}
J_\mathbf{P}\phi_{klm}=i\phi_{klm}\,,\quad J_\mathbf{P}\bar{\phi}_{klm}=-i\bar{\phi}_{klm}\,.\label{J_phi}
\end{equation}
As we will see in the next section, these choices represent  all the
possibilities for the  $SO(4)$-invariant complex structures in $\mathcal{S}$.
Here we discuss the freedom in the choice of the
normalized set $\{y_k\}$. To this end, let us fix once and for all a
family $\{y_{0k}=u_{0k}+iv_{0k}\}$ verifying the normalization
condition
$$\bar{y}_{0k}(t)\dot{y}_{0k}(t)-\dot{\bar{y}}_{0k}(t)y_{0k}(t)=\frac{i}{f^2(t)\cosh^{3}(t)}\,.$$
For example by considering the pairs $(u_{0k},v_{0k})$ given by
(\ref{u0v0}). Then, any  other normalized set $\{y_k\}$ can be
expressed in terms of $u_{0k}$ and $v_{0k}$ as
\begin{equation}
y_{k}(t)=\alpha_{k}u_{0k}(t)+\beta_{k}v_{0k}(t)
+i(\gamma_{k}u_{0k}(t)+\delta_{k}v_{0k}(t))\,,
\end{equation}
where the real coefficients
$\alpha_{k},\beta_{k},\gamma_{k},\delta_{k}$ must
satisfy
\begin{equation}
\mathrm{det}\left(\begin{array}{cc} \alpha_{k} & \beta_{k}\\
\gamma_{k} & \delta_{k}\end{array}\right)=1\Leftrightarrow \left(\begin{array}{cc} \alpha_{k} & \beta_{k}\\
\gamma_{k} & \delta_{k}\end{array}\right)\in
SL(2,\mathbb{R})\,,\,\,\forall\,k\in\mathbb{N}_{0}\,.
\end{equation}
It is well known that $SL(2,\mathbb{R})$ is bijective (as a set) to
$\mathbb{S}^{1}\times\mathbb{R}^{2}$, in the sense that  any element
of $SL(2,\mathbb{R})$ can be uniquely decomposed as a product of a
rotation and an upper triangular matrix with unit determinant
\begin{equation}
SL(2,\mathbb{R})\ni\left(\begin{array}{cc} \alpha_{k} & \beta_{k}\\
\gamma_{k} & \delta_{k}\end{array}\right)=\left(\begin{array}{cc} \cos\theta_{k} & -\sin\theta_{k}\\
\sin\theta_{k} & \cos\theta_{k}\end{array}\right)\left(\begin{array}{cc} \rho_{k} & \nu_{k}\\
0 & \rho_{k}^{-1}\end{array}\right),\label{polar}
\end{equation}
for a unique choice of $\rho_{k}>0$, $\nu_{k}\in\mathbb{R}$,
$\theta_{k}\in[0,2\pi)$. Hence, different choices of the triplet
$(\rho_k,\nu_k,\theta_k)$ will, in principle, correspond to different complex structures
on the space of solutions $\mathcal{S}$, defined through (\ref{J_phi}) where
\begin{eqnarray}
y_k(t)&=&\rho_k\cos\theta_k u_{0k}(t)+(\nu_k\cos\theta_k-\rho^{-1}_k\sin\theta_k)v_{0k}(t)\label{y_k_general}\\
&+&i\Big(\rho_k\sin\theta_k u_{0k}(t)+(\nu_k\sin\theta_k+\rho^{-1}_k\cos\theta_k)v_{0k}(t)\Big)\,.\nonumber
\end{eqnarray}
However, it is easy to show \cite{G.:2007rd} that two different
choices of the form $(\rho_k,\nu_k,\theta_k)$ and
$(\rho_k,\nu_k,\tilde{\theta}_k)$ give rise to the same complex
structure. Then, in the following, we will omit the angular part of
$(\rho_k,\nu_k,\theta_k)$ by choosing $\theta_k=0$ in (\ref{polar}).

The complex structures defined trough $(\rho_k,\nu_k)$ will, in
general, yield irreducible unitarily nonequivalent representations
of the canonical commutation relations. The existence of many
non-unitarily equivalent quantization is a well known property of
any QFT in a generic curved space-time, and can be considered as a
serious drawback to the formulation of the theory. Obviously, this
is not the case for a system of finite number of degrees of freedom,
where the Stone-von Neumann theorem can be applied \cite{Prugovecki}: for any Lagrangian
subspace $\mathbf{P}$ one obtains a quantum theory unitarily
equivalent to the usual tensor product construction. Also, for the
case of a massless scalar field evolving in a fixed
\emph{stationary} space-time, there exists a preferred choice of
Lagrangian subspace by virtue of the time translation symmetry
\cite{Wald}. In our case, in absence of this symmetry, no natural,
preferred election of $\mathbf{P}$ is available; in other words, due
to the time-dependence of the Hamiltonian (\ref{hamiltonian}) the
solutions of $\mathcal{S}$ do not oscillate harmonically, and thus
it is not possible to uniquely define subspaces of positive and
negative frequency solutions.

\subsection{$SO(4)$-invariant complex structures}{\label{Invariant complex structures}}

Consider some specific basis $\{\phi_{0klm}\}$ of
$\mathcal{H}_{\mathbf{P}}$ and the corresponding splitting of the
complexified solution space
$\mathcal{S}_{\mathbb{C}}=\mathbf{P}_0\oplus\bar{\mathbf{P}}_0$. For
practical purposes, let us denote
\begin{eqnarray*}
\mathbf{P}_{1}&:=&\mathbf{P}_{0}=\bigoplus_{k=0}^{\infty}\mathbf{P}_{1}^{k}\,,\,\,\,\mathbf{P}^k_1:=\mathrm{span}\{y_{0k}\}\otimes\mathrm{span}\{Y_{k
l
m}\,|\, l\in \{0,1,\dots, k\},\, m\in\{-l,\dots,l\}\}, \\
\mathbf{P}_{2}&:=&\bar{\mathbf{P}}_{0}=\bigoplus_{k=0}^{\infty}\mathbf{P}_{2}^{k}\,,\,\,\,\mathbf{P}_2^k:=\mathrm{span}\{\bar{y}_{0k}\}\otimes\mathrm{span}\{Y_{kl
m}\,|\, l\in \{0,1,\dots, k\},\, m\in\{-l,\dots,l\}\}.
\end{eqnarray*}
The elements $\phi_a\in \mathbf{P}_a$, $a=1,2$, are complex
functions $\phi_a(t,s)$ defined on $\mathbb{R}\times \mathbb{S}^3$.
The natural representation $D_a$ of $SO(4)$ in $\mathbf{P}_a$ is then
defined by $(D_a(g)\phi)(t,s)=\phi(t,g^{-1}\cdot s)$, where
$g^{-1}\cdot s$ denotes the action of the rotation $g^{-1}\in SO(4)$
on the point $s\in \mathbb{S}^3$. This allows us to write the
representation of $SO(4)$ in
$\mathcal{S}_{\mathbb{C}}=\mathbf{P}_1\oplus \mathbf{P}_2$ in matrix
form
$$
D(g)=\left(\begin{array}{cc} D_1(g)&0\\0&D_2(g)
\end{array}\right),\quad g\in SO(4)\,,
$$
in terms of the representations $(D_a,\mathbf{P}_a)$. In particular,
we are interested in characterizing the complex structures $J$
defined on the solution space $\mathcal{S}$ invariant under the
$SO(4)$ symmetry of the round 3-sphere $\mathbb{S}^{3}$. Using the
previous notation, this implies
\begin{equation}\label{inv}
D(g)J=J D(g)\Leftrightarrow \left(\begin{array}{cc}
J_{11}D_1(g)&J_{12}D_2(g)\\J_{21}D_1(g)&J_{22}D_2(g)
\end{array}\right)=\left(\begin{array}{cc}
D_1(g)J_{11}&D_1(g)J_{12}\\D_2(g)J_{21}&D_2(g)J_{22}
\end{array}\right),\,\,\, \forall g\in SO(4)\,,
\end{equation}
in terms of $\mathbb{C}$-linear maps
$J_{ab}:\mathbf{P}_{b}\rightarrow\mathbf{P}_{a}$, $a,b\in\{1,2\}$.
The same arguments used in \cite{G.:2007rd} with the aim of
identifying the $SO(3)$-invariant complex structures for the Gowdy
$\mathbb{S}^{1}\times\mathbb{S}^{2}$ and $\mathbb{S}^{3}$ models
show now that the general form of a $\mathbb{R}$-linear operator $J$
satisfying (\ref{inv}) is given by
$$
J=\bigoplus_{k=0}^{\infty}\left(
\begin{array}{cc}
\jmath_{11}^k I_{11}^k&\jmath_{12}^k I_{12}^k\\
\bar{\jmath}_{12}^k I_{21}^k&\bar{\jmath}_{11}^k I_{22}^k
\end{array}
\right),
$$
where $I_{aa}^k$ denotes the identity operator in $\mathbf{P}_a^k$
and the linear operators
$I_{ab}^k:\mathbf{P}_b^k\rightarrow\mathbf{P}_a^k$ act according to
$I_{12}^k(\bar{y}_{0k}\otimes v)=y_{0k}\otimes v$ and
$I_{21}^k(y_{0k}\otimes v)=\bar{y}_{0k}\otimes v$. Finally, due to
the extra condition $J^2=-\mathrm{Id}_{\mathbb{S}}$, defining $J$ as
a complex structure, the complex coefficients $\jmath_{11}^k$ and
$\jmath_{12}^k$ should verify
\begin{equation}
|\jmath_{11}^k|\,^2-|\jmath_{12}^k|\,^2=1,\quad \jmath_{11 }^k\in i
\mathbb{R}\smallsetminus\{0\}\,,\quad \jmath_{12}^k\in \mathbb{C}\,.
\label{conds}
\end{equation}
According to this result, it suffices to take $\jmath^{k}_{11}=i$ and
$\jmath^{k}_{12}=0$, $\forall\,k\in\mathbb{N}_{0}$, to conclude that
all the complex structures $J_{\mathbf{P}}$ defined in  section \ref{Complex
structures and mode decomposition} are $SO(4)$-invariant. It is also
clear that any $SO(4)$-invariant complex structure, characterized by
the pairs $(\jmath^{k}_{11},\jmath^{k}_{12})$ verifying
(\ref{conds}), has an associated Lagrangian subspace $\mathbf{P}$
defined by the set $\{y_{k}=\rho_k
u_{0k}+(\nu_k+i\rho_k^{-1})v_{0k}\}$, with the coefficients
$(\rho_k,\nu_k)$ related to $(\jmath^{k}_{11},\jmath^{k}_{12})$ by
\begin{eqnarray*}
\jmath_{11}^k=\frac{i}{2}(\nu_k^2+\rho^{-2}_k+\rho^2_k)\,,\quad
\jmath_{12}^k=
-\rho_k\nu_k+\frac{i}{2}(\nu_k^2+\rho^{-2}_k-\rho^2_k)\,.
\end{eqnarray*}

\subsection{Canonical quantum field operators}

 The canonical field operators  associated
to a given time $t\in\mathbb{R}$ are defined as operator-valuated
distributions on $\mathbb{S}^{3}$
\begin{eqnarray}
\hat{Q}(t,s)&=&\sum_{k=0}^{\infty}\sum_{l=0}^{k}\sum_{m=-l}^{l}\left(y_{k}(t)Y_{klm}(t)\hat{a}_{klm}+\bar{y}_{k}(t)\bar{Y}_{klm}(s)\hat{a}_{klm}^{\dag}\right),\label{opQ}\\
\hat{P}(t,s)&=&\sum_{k=0}^{\infty}\sum_{l=0}^{k}\sum_{m=-l}^{l}\Big(\big[f^2(t)\cosh^3t\dot{y}_{k}(t)+f(t)\dot{f}(t)\cosh^3t y_k(t)\big]Y_{klm}(t)\hat{a}_{klm}\label{opP}\\
& & \hspace{2.3cm}+\big[f^2(t)\cosh^3t\dot{\bar{y}}_{k}(t)+f(t)\dot{f}(t)\cosh^3t \bar{y}_k(t)\big]\bar{Y}_{klm}(s)\hat{a}_{klm}^{\dag}\Big).\nonumber
\end{eqnarray}
In practice, these expressions can be obtained by formally promoting
the Fourier coefficients in (\ref{isomor}) to the creation and
annihilation operators --$\hat{a}_{klm}^{\dag}$ and $\hat{a}_{klm}$,
respectively-- associated to the basic vectors
$\phi_{klm}\in\mathcal{H}_\mathbf{P}$. Given any pair of smooth
real-valued functions on the 3-sphere $g_1$, $g_2\in
C^{\infty}(\mathbb{S}^{3};\mathbb{R})$, these distributions define
canonical field operators $(\hat{Q}[g_1](t),\hat{P}[g_2](t))$
obtained by multiplying the formal expressions (\ref{opQ}) and
(\ref{opP}) by $g_1$ and $g_2$, respectively, and integrating over
the round 3-sphere $\mathbb{S}^{3}$. In the next section we will
study the behavior of these canonical operators in $t$. In
particular we will consider if it is possible to choose the
conformal factor $f$ and the complex structure, defined through
$\{y_k\}$, in such a way that the functional dependence in $t$  of
$(\hat{Q}(t,s),\hat{P}(t,s))$ can be obtained by the action of an
unitary operator in the Fock space. As we will see, this is not the
case.

\section{Classical dynamics and unitary quantum evolution}{\label{Classical dynamics and unitary quantum evolution}}

Classical time evolution  from the embedding
$\imath_{t_0}(\mathbb{S}^{3})\subset \mathbb{R}\times \mathbb{S}^3$
to $\imath_{t_1}(\mathbb{S}^{3})\subset\mathbb{R}\times
\mathbb{S}^3$ is implemented on the canonical phase space $\Upsilon$
by the symplectic transformation
$\tau_{(t_0,t_1)}:\Upsilon\rightarrow\Upsilon$ defined as
$$\tau_{(t_0,t_1)}:=\mathfrak{J}_{t_1}\circ \mathfrak{J}_{t_0}^{-1}$$
in terms of the symplectic maps $\mathfrak{J}_t$ introduced in
(\ref{isomor}) and their inverses
$\mathfrak{J}_t^{-1}:\Upsilon\rightarrow \Gamma$,
$\zeta=\mathfrak{J}_t^{-1}(Q,P)$. The maps $\mathfrak{J}_t^{-1}$ can
be easily  computed in terms of the Fourier coefficients
(\ref{modes}) of $\zeta$ as
\begin{equation}
a_{klm}(t)=i(f(t)\dot{\bar{y}}_k(t)+\dot{f}(t)\bar{y}_k(t))f(t)\cosh^3
t \int_{\mathbb{S}^3} |\gamma|^{1/2}
Y_{klm}Q-i\bar{y}_k(t)\int_{\mathbb{S}^3} |\gamma|^{1/2}
Y_{klm}P\,.\label{inversas}
\end{equation}
The operator $\tau_{(t_0,t_1)}$ acts as follows: (i) first, it takes
initial Cauchy data on $\imath_{t_0}(\mathbb{S}^{3})$, (ii) evolves
them to the corresponding solution in $\mathcal{S}$, and (iii) finds
the Cauchy data induced by this solution on
$\imath_{t_1}(\mathbb{S}^{3})$. On the other hand, time evolution
can also be viewed as a symplectic transformation on the covariant
phase space, $\mathcal{T}_{(t_0,t_1)}:\Gamma\rightarrow\Gamma$,
defined by
\begin{equation}
\mathcal{T}_{(t_0,t_1)}:=\mathfrak{J}_{t_1}^{-1}\circ\tau_{(t_0,t_1)}\circ\mathfrak{J}_{t_1}
=\mathfrak{J}_{t_0}^{-1}\circ\mathfrak{J}_{t_1}\,,
\label{eee}
\end{equation}
that (i) takes a solution of $\mathcal{S}$, (ii) finds the induced
Cauchy data on $\imath_{t_1}(\mathbb{S}^{3})$, and (iii) takes that data
as initial data on $\imath_{t_0}(\mathbb{S}^{3})$, finding finally the
corresponding solution. In our case, combining (\ref{isomor}) and (\ref{inversas}), it is straightforward to check that, whenever
$$
\zeta=\sum_{k=0}^{\infty}\sum_{l=0}^{k}\sum_{m=-l}^{l}\Big(a_{klm}\phi_{klm}
+\bar{a}_{klm}\bar{\phi}_{klm}\Big)\,,
$$
the action of the operator $\mathcal{T}_{(t_0,t_1)}$ is given by
$$
\mathcal{T}_{(t_0,t_1)}\zeta =\sum_{k=0}^{\infty}\sum_{l=0}^{k}\sum_{m=-l}^{l}\Big(\mathfrak{a}_{klm}(t_0,t_1)\phi_{klm}
+\bar{\mathfrak{a}}_{klm}(t_0,t_1)\bar{\phi}_{klm}\Big)\,,
$$
where
\begin{eqnarray}
\mathfrak{a}_{klm}(t_0,t_1)&:=&
i\left(\frac{f(t_0)}{f(t_1)}\cosh^3(t_0)\dot{\bar{z}}_k(t_0)z_k(t_1)
-\frac{f(t_1)}{f(t_0)}\cosh^3(t_1)\dot{z}_k(t_1)\bar{z}_k(t_0)\right)\, a_{klm}\\
&+& i\left(
\frac{f(t_0)}{f(t_1)}\cosh^3(t_0)\dot{\bar{z}}_k(t_0)\bar{z}_k(t_1)
-\frac{f(t_1)}{f(t_0)}\cosh^3(t_1)\dot{\bar{z}}_k(t_1)\bar{z}_k(t_0)\right)\,\bar{a}_{kml}\,.\nonumber
\end{eqnarray}
Here we have introduced the functions
$$
z_k(t):=f(t)y_k(t)
$$
that will allow us to use a more economical notation in what
follows.

\subsection{Unitary implementation of quantum dynamics}{\label{Quantum unitary implementability}}

Let us consider now the quantum counterpart of the classical time
evolution. Taking a $SO(4)$-invariant complex structure
$J_{\mathbf{P}}$  and
using the theory of unitary implementation of symplectic
transformations \cite{Shale,Wald}, the time evolution
$\mathcal{T}_{(t_0,t_1)}$ defined in (\ref{eee}) is unitarily implementable on the Fock
space $\mathcal{F}_{s}(\mathcal{H}_{P})$, i.e., there exists a unitary operator
$\hat{U}(t_0,t_1):\mathcal{F}_{s}(\mathcal{H}_{\mathcal{P}})\rightarrow\mathcal{F}_{s}(\mathcal{H}_{P})$
such that
\begin{equation}
\hat{U}^{-1}(t_0,t_1)\hat{Q}(t_0,s)\hat{U}(t_0,t_1)
=\hat{Q}(t_1,s)\,,\,\,\,\hat{U}^{-1}(t_0,t_1)\hat{P}(t_0,s)\hat{U}(t_0,t_1)=\hat{P}(t_1,s)\,,
\end{equation}
if and only if $J_{\mathbf{P}}-\mathcal{T}_{(t_0,t_1)}^{-1}\circ
J_{\mathbf{P}}\circ\mathcal{T}_{(t_0,t_1)}$ is Hilbert-Schmidt. This is equivalent to demand that
\begin{equation}
\sum_{k=0}^{\infty}(k+1)^{2}|\beta_k(t_0,t_1|z_k,f)|^{2}<\infty\,,\label{kkk}
\end{equation}
where
\begin{equation}
\beta_k(t_0,t_1|z_k,f):=\frac{f(t_0)}{f(t_1)}\cosh^3(t_0)\dot{z}_k(t_0)z_k(t_1)
-\frac{f(t_1)}{f(t_0)}\cosh^3(t_1)\dot{z}_k(t_1)z_k(t_0)\,.
\end{equation}
Notice that the square summability of the series (\ref{kkk}) depends only on its ultraviolet behavior. In particular, the zero mode corresponding to
$k=0$ plays no role in this context. As we pointed out in
subsection \ref{Invariant complex structures}, the $SO(4)$-invariant
complex structures differ from each other just in the pair
$(\rho_k,\nu_k)$ which appears inside of each $z_k=fy_k$. Then we
will consider the complex structures induced by choosing $z_k=fy_k$
of the form
$$z_{k}(t)=\rho_kf(t)u_{0k}(t)+(\nu_k+i\rho_k^{-1})f(t)v_{0k}(t)
=\rho_k\tilde{u}_{0k}(t)+(\nu_k+i\rho_k^{-1})\tilde{v}_{0k}(t)$$
with
\begin{eqnarray}
\tilde{u}_{0k}(t)&:=&f(t)u_{0k}(t)=\frac{2^{(k+1)}}{\sqrt{3}}\cosh^{k}(t)\mathrm{e}^{-kt}
\,{_{2}}F_{1}(k+3/2,k;-1/2;-\mathrm{e}^{-2t})\,,\\
\tilde{v}_{0k}(t)&:=&f(t)v_{0k}(t)=-\frac{2^{(k+1)}}{\sqrt{3}}\cosh^{k}(t)\mathrm{e}^{-(k+3)t}
\,{_{2}}F_{1}(k+3/2,k+3;5/2;-\mathrm{e}^{-2t})\,,\\
\dot{\tilde{u}}_{0k}(t)&=&-\frac{2^{(k+1)}}{\sqrt{3}}k\cosh^{k-1}(t)\mathrm{e}^{-(k+2)t}\bigg(\mathrm{e}^{t}
\,{_2}F_{1}(k,k+3/2;-1/2;-\mathrm{e}^{-2t})\nonumber\\
&&+2(2k+3)\cosh(t){_2}F_{1}(k+1,k+5/2;1/2;-\mathrm{e}^{-2t})\bigg)\,,\\
\dot{\tilde{v}}_{0k}(t)&=&\frac{2^{k}}{5\sqrt{3}}\cosh^{k-1}(t)\mathrm{e}^{-(k+5)t}
\bigg(5(3+2k+3\mathrm{e}^{2t})\mathrm{e}^{t}
\,{_2}F_{1}(k+3/2,k+3;5/2;-\mathrm{e}^{-2t})\nonumber\\
&&-4(k+3)(2k+3)\cosh(t)\,{_2}F_{1}(k+5/2,k+4;7/2;-\mathrm{e}^{-2t})\bigg)\,.
\end{eqnarray}
Taking into account the asymptotic behaviors (valid for $\lambda\rightarrow+\infty$, $\alpha,\beta\in\mathbb{R}$, $0<z<1$)
\begin{eqnarray*}
{_2}F_{1}\bigg(\lambda+\alpha,\lambda+\beta;\frac{5}{2};-z\bigg)&\sim&-\frac{3}{4z\lambda^{2}}(1+z)^{\frac{5}{4}-\lambda-\frac{\alpha+\beta}{2}}\cos\Theta(\lambda,\alpha+\beta-2;z)\,,\\
{_2}F_{1}\bigg(\lambda+\alpha,\lambda+\beta;-\frac{1}{2};-z\bigg)&\sim&2\lambda\sqrt{z}(1+z)^{-\frac{1}{4}-\lambda-\frac{\alpha+\beta}{2}}\sin\Theta(\lambda,\alpha+\beta+1;z)\,,\\
{_2}F_{1}\bigg(\lambda+\alpha,\lambda+\beta;\frac{1}{2};-z\bigg)&\sim&(1+z)^{\frac{1}{4}-\lambda-\frac{\alpha+\beta}{2}}\cos\Theta(\lambda,\alpha+\beta;z)\,,\\
{_2}F_{1}\bigg(\lambda+\alpha,\lambda+\beta;\frac{7}{2};-z\bigg)&\sim&-\frac{15}{8}z^{-3/2}\lambda^{-3}(1+z)^{\frac{7}{4}-\lambda-\frac{\alpha+\beta}{2}}\sin\Theta(\lambda,\alpha\hspace{-0.05cm}
+\hspace{-0.05cm}\beta\hspace{-0.05cm}-\hspace{-0.08cm}3;z)\,,
\end{eqnarray*}
with
\begin{equation}
\Theta(\lambda,\gamma;z):=(2\lambda+\gamma)\arctan\sqrt{z}-\arctan\frac{\sqrt{z}}{1+\sqrt{1+z}}\,,
\end{equation}
we get that, for $k\rightarrow+\infty$,
\begin{eqnarray*}
\tilde{u}_{0k}(t)&\sim&\frac{2}{\sqrt{3}}\frac{k}{\cosh(t)}\sin\Theta(k,5/2;\mathrm{e}^{-2t})\,,\\
\tilde{v}_{0k}(t)&\sim&\frac{\sqrt{3}}{4k^{2}}\frac{1}{\cosh(t)}\cos\Theta(k,5/2;\mathrm{e}^{-2t})\,,\\
\dot{\tilde{u}}_{0k}(t)&\sim&-\frac{2k^{2}}{\sqrt{3}}\frac{\mathrm{e}^{-2t}}{\cosh(t)}\bigg(2(1+\mathrm{e}^{-2t})^{-1}\sin\Theta(k,5/2;\mathrm{e}^{-2t})\nonumber\\
&&+4\cosh(t)(1+\mathrm{e}^{-2t})^{-\frac{3}{2}}\cos\Theta(k,7/2;\mathrm{e}^{-2t})\bigg)\,,\\
\dot{\tilde{v}}_{0k}(t)&\sim&-\frac{\sqrt{3}}{2k}\frac{\mathrm{e}^{-2t}}{\cosh(t)}\bigg((1+\mathrm{e}^{-2t})^{-1}\cos\Theta(k,5/2;\mathrm{e}^{-2t})\nonumber\\
&&-2\cosh(t)(1+\mathrm{e}^{-2t})^{-\frac{3}{2}}\sin\Theta(k,7/2;\mathrm{e}^{-2t})\bigg)\,.
\end{eqnarray*}
These asymptotic expansions can be obtained by using the steepest descent method in a multi-variable integral representation of the hypergeometric function ${_2}F_{1}$ \cite{Barbero}. Since
\begin{eqnarray*}
\mathrm{Im}[\beta_k(t_0,t_1|z_k,f)]&=&\frac{f(t_0)}{f(t_1)}\cosh^{3}(t_0)\Bigg(\dot{\tilde{u}}_{0k}(t_0)\tilde{v}_{0k}(t_1)+
\dot{\tilde{v}}_{0k}(t_0)\tilde{u}_{0k}(t_1)
+2\nu_{k}\rho^{-1}_{k}\dot{\tilde{v}}_{0k}(t_0)\tilde{v}_{0k}(t_1)\Bigg)\nonumber\\
&-&\frac{f(t_1)}{f(t_0)}\cosh^{3}(t_1)\Bigg(\dot{\tilde{u}}_{0k}(t_1)\tilde{v}_{0k}(t_0)+
\dot{\tilde{v}}_{0k}(t_1)\tilde{u}_{0k}(t_0)
+2\nu_{k}\rho^{-1}_{k}\dot{\tilde{v}}_{0k}(t_1)\tilde{v}_{0k}(t_0)\Bigg),
\end{eqnarray*}
it is straightforward to show that, irrespective of the function $f>0$  and the
complex structure parameterized by $(\rho_k,\nu_k)$,
\begin{equation}
\lim_{k\rightarrow\infty}\mathrm{Im}[\beta_k(t_0,t_1|z_k,f)]\neq 0\,.
\end{equation}
Hence, it is \textit{not} possible to find a conformal factor $f$
and a complex structure $J_\mathbf{P}$ such that the dynamics can be unitarily implemented on
$\mathcal{F}_{s}(\mathcal{H}_{\mathbf{P}})$.

\section{Conclusions and comments}{\label{Conclusions and comments}}

Some comments are in order now. First of all, by specializing the
previous result to the choice  $f(t)=1$, we have shown in section
\ref{Quantum unitary implementability} that it is impossible to find
a  Fock quantization of a minimally coupled massless scalar field $\phi$ propagating
in de Sitter space-time such that the quantum dynamics can be
unitarily implemented. The only constraint that our Fock
quantizations should satisfy is that they must be derived from a
$SO(4)$-invariant complex structure on the space of classical
solutions. The main result of the paper is that, in contrast with
the Gowdy models, in the case of massless scalars in de Sitter
space-time it is not possible to attain the quantum unitarity of the
evolution by considering a spatially homogeneous field redefinition
of the form $\zeta(t,s)=f^{-1}(t)\phi(t,s)$.

\bigskip

It is important to notice that for the $\mathbb{S}^1\times
\mathbb{S}^2$ and $\mathbb{S}^3$ Gowdy models the conformal
transformation that leads to a well behaved quantum evolution is the
one that relates the background Gowdy metrics to the Einstein static
universe \cite{BarberoG.:2007,G.:2007rd}. In the case of the de
Sitter space, the transformation that connects the de Sitter metric
to the  Einstein static universe leads us to consider
$f(t)=1/\cosh(t)$. Now the scalar field $\zeta$ propagates in the
Einstein static background universe but verifying a Klein-Gordon
equation with a tachyonic time-dependent mass term. Explicitly the
field $\zeta$ satisfies
\begin{equation}
\ddot{\zeta}+\tanh(t)\dot{\zeta}-\sech\!^2\,t
\big(\cosh(2t)+\Delta_{\mathbb{S}^3}\big)\zeta=0\,.\label{eqcosh}
\end{equation}
Then, if we use the time coordinate
$T=2\arctan\left(\tanh(t/2)\right)$  defined in (\ref{Ttime}) in
order to see the relation  between the de Sitter and Einstein
metrics, we get
$$
\partial_T^2\zeta-\big(\cosh(2t)+\Delta_{\mathbb{S}^3}\big)\zeta=0\,.
$$
It is clear that the time-dependent `mass term' $\cosh(2t)$ has the
`wrong sign'. Due to this fact, the modes (\ref{u0v0}) satisfy the
harmonic oscillator equation with a time depende square frequency
$k(k+2)-\cosh(2t)$ whose sign, irrespectively
of $k$, becomes negative when $t\rightarrow
\pm\infty$ (or equivalently $T\rightarrow \pm\pi/2$) . This introduces a non oscillatory
behavior of these modes when $t\rightarrow \pm\infty$   that, at the
end of the day, is one of the factors responsable of the failure of
the unitarity condition.

It could appear at first sight that the failure of unitarity is also
related to the presence of a first derivative term in the wave
equation (\ref{eqcosh}). This conclusion would be motivated by the
Gowdy cases where the time variable usually used is such that the
metric appears in the form
$\mathring{g}_{ab}=-(\mathrm{d}t)_a(\mathrm{d}t)_b+\gamma^\Sigma_{ab}$.
In this way, the field $\zeta$ for which evolution is well defined
verifies an equation
$\mathring{g}^{ab}\mathring{\nabla}_a\mathring{\nabla}_b\zeta=-\ddot{\zeta}+\Delta_\Sigma
\zeta=\frac{1}{4}(1+\csc^{2}t)\zeta$ without linear terms in
$\dot{\zeta}$. On the contrary, for the 4-dimensional de Sitter
case, the metric $\mathring{g}_{ab}$ has the form
$-(\mathrm{d}T)_a(\mathrm{d}T)_b+\gamma_{ab}=-\dot{T}^2(\mathrm{d}t)_a(\mathrm{d}t)_b+\gamma_{ab}$,
so that the field $\zeta$ verifies the equation (\ref{eqcosh}) that
involves a linear term in $\dot{\zeta}$. However, when the time
coordinate $T=T(t)$ is used instead of $t$, the $\partial_T\zeta$-term
disappear from the equation of motion. In conclusion, our results
regarding unitarity do not depend on the way the evolution
is parameterized.

\bigskip

Owing to the fact that the dynamical evolution is not unitarily
implementable, one is faced with the non-existence of the usual
Schr\"{o}dinger picture. As far as models with a finite number of
degrees of freedom is concerned, this situation is normally related
to the lack of a suitable probabilistic interpretation of the
theory. However, this is not necessarily the case for systems of
infinitely many degrees of freedom, as it was pointed out for
instance in \cite{Torre:2002xt} for the case of the Gowdy
$\mathbb{T}^3$ model. Hence, the impossibility of implementing the
dynamics in a unitary way is not an obstacle to consider this type
of models as physically relevant. Indeed, concerning the de Sitter
case, it is possible to address some physical sensible questions
such as the cosmological constant problem \cite{Mottola:1984ar}
despite the absence of unitary dynamics.

In conclusion, the results discussed in this paper show that the
method successfully used to obtain a unitary dynamics in the Fock
quantization of the Gowdy models --based on a suitable redefinition
of the scalar fields suggested by the conformal factor of the
background metric-- is not of general validity, not even in the context of highly symmetric space-time backgrounds.

\begin{acknowledgments}
The authors want to thank Fernando Barbero
for many interesting discussions. Daniel G\'omez Vergel acknowledges the support of
the Spanish Research Council (CSIC) through a I3P research
assistantship. This work is also supported by the Spanish MEC under
the research grant FIS2005-05736-C03-02.

\end{acknowledgments}

\end{document}